# The Smithsonian/NASA Astrophysics Data System (ADS) Decennial Report

**Michael J. Kurtz**
**Alberto Accomazzi**
**Stephen S. Murray**
*Harvard-Smithsonian Center for Astrophysics*
*{mkurtz,aaccomazzi,smurray}@cfa.harvard.edu*

## Summary

Eight years after the ADS first appeared the last decadal survey wrote: "NASA's initiative for the Astrophysics Data System has vastly increased the accessibility of the scientific literature for astronomers. NASA deserves credit for this valuable initiative and is urged to continue it." Here we summarize some of the changes concerning the ADS which have occurred in the past ten years, and we describe the current status of the ADS. We then point out two areas where the ADS is building an improved capability which could benefit from a policy statement of support in the ASTRO2010 report. These are: The Semantic Interlinking of Astronomy Observations and Datasets and The Indexing of the Full Text of Astronomy Research Publications.

## Highlights in the Early History of the ADS

The ADS was conceived in 1988 and first came on-line in 1992 (Kurtz, et al 1993).  In 1993 we made links to what has now become the Vizier project at the CDS in Strasbourg, and by the end of that year were able to make joint queries with the SIMBAD database in Strasbourg, thus permitting queries combining objects and keywords such as "show me papers which are about M87 and which contain the phrase *globular cluster* in the abstract."

In November 1993 the World Wide Web became mature enough to support fielded queries. A few months later the ADS became a web-based service.  In 1994 we began to digitize the (paper) back-issues of astronomy journals.  When the ApJ Letters came on-line in mid 1995, its electronic articles linked to ADS records from their reference sections.  By that time, the ADS had already put more than ten years of back-issues on-line, up to the first electronic issue of the journal.

In 1996 the ADS began linking datasets in archives with journal articles (first with NCSA). In the same year the ADS began maintaining a citation index for the astronomy literature, thanks to the financial support of the AAS.  In 1997 the ADS began indexing the *astro-ph* preprints, from what is now known as the arXiv. In the summer of 1997 the number of full text downloads of journal articles mediated by the ADS surpassed the total number of astronomy journal articles read in all the print libraries of the world.  This was six months after ApJ and A&A officially went electronic, and six months before AJ and MNRAS did.

At the end of the decade the Decadal Survey said: "NASA's initiative for the Astrophysics Data System has vastly increased the accessibility of the scientific literature for astronomers. NASA deserves credit for this valuable initiative and is urged to continue it."  A discussion of the early history of the ADS can be found in Kurtz, et al (2000), with an update (as of 2003) in Kurtz, et al (2005).

## Highlights of the Last Ten Years of the ADS

Over the last ten years the ADS has substantially increased its holdings and its reach. Some of the most important highlights and improvements have been:

- bibliographic records in the ADS databases increased by a factor of 5, to 7.5M
- pages of full-text in the ADS increased by a factor of 3, to 3.8M
- number of citations in the citation index increased by a factor of 10, to 33M
- number of links to tables or archival data files increased by a factor of 10, to 60K
- data items downloaded by scientists increased by a factor of 10, to 23M per year
- data items downloaded by the general public (via Google, Google Scholar, etc.) increased from zero to 24M per year
- the number of unique monthly users has often exceeded 1M
- implementation of full-text searching of the historical literature
- introduction of the *myADS Notification Service*, which now has more than 6K subscribers
- introduction, in collaboration with the arXiv, of the *myADS-arXiv* notification service which may be viewed as an individually customized open-access virtual journal
- creation the ADS Physics/Astronomy Education Service, in collaboration with the ComPadre project
- creation the Private Library feature, now used by more than 12K individuals

## Current Status of the ADS

The ADS is quadrennially reviewed as part of the NASA Senior Review of Astrophysics Data Archives.  Historically, the ADS has always received positive reviews.  The most recent review, which took place in 2008, ranked it number one.  For the past decade the ADS (combined with the SIMBAD USA project) has been staffed at 6.25 FTE, although for the past 20 months we have operated one person short.  Following the Senior Review our budget has been increased to allow a staff of 7.5 FTE, and we expect to be fully staffed later this year.  The ADS management believes that the ADS budget is adequate to support its responsibilities.  We thank NASA for its ongoing support and confidence in us.

A prerequisite for any project providing an important part of infrastructure is its long term stability.  If outside organizations are to rely on services that we provide they must have the confidence to expend substantial resources to do so. In the terms of the NSF Cyberinfrastructure Council's "Cyberinfrastructure Vision for 21st Century Discovery" (NSF 2007), the ADS is a *Reference Collection*: it is a core facility of astrophysics research.  Our highest priority for the future is to implement policies and practices which ensure the long term stability and growth of the ADS. To this end we are replacing our aging, custom-built software with a modern, open source system.  The ADS was one of the first internet search systems, predating AltaVista by two years, PubMed by four years and Google by six years. As one of the early adopters of web-based technologies, the ADS did not have the opportunity to take advantage of the technology and best practices now widely in use, and its IT infrastructure, which was developed over the last 15 years, has grown obsolete and difficult to maintain.  We are currently collaborating with the High Energy Physics community in an effort to adopt and extend CERN's Invenio system, an open source project which is also being adopted by the Spires system run by SLAC, DESY, and Fermilab.  We are also taking steps to secure the long term stability and integrity of our hardware and software platforms by implementing a disaster recovery plan which makes use of a full backup system at a site in Herndon, VA run by the Smithsonian Institution.  This will allow

ADS to continue operating in the event of a catastrophic failure at our Cambridge, MA site.

## Two Current ADS Projects

The ADS has always been a thought leader in scientific search systems. We expect to continue our leadership role, and are working in a number of directions towards that end. Two of the ADS's most important ongoing projects could benefit greatly from statements of support from the committee. Although we have the cooperation of most of the largest organizations, the systems we are building cannot be fully successful without the participation of *all* major players. We note expressly that the support we are seeking is at the policy rather than funding level, which is adequate to the tasks at hand.

### 1. The Semantic Interlinking of Astronomy Observations and Datasets

Astronomy is an observational science. Our theories of the behavior of the universe and the objects in it are inspired by observations of that behavior, and are supported (or rejected) by observations, as well. Within the past twenty years systems of dense, interlinked data have become common and expected, fundamentally changing the way people think and services operate. It is now expected, for example, that from a listing of a movie at our local theater one can find a synopsis, reviews, a cast list, a list of all the director's films, etc. One expects to be able to buy tickets, as well.

Astronomy was one of the first disciplines to benefit from the early developments of the web-based technologies enabling cross-linking of resources across archives (Accomazzi et al, 1994). Sixteen years ago, thanks to a collaboration between the ADS and the CDS, it became possible to go from a list of articles to the abstract of an article to a list of astronomical objects described in that article to a set of measurements on one of those objects. Thirteen years ago, again thanks to a collaboration between the ADS and several major data centers, including HEASARC, MAST, ESO, and Chandra, it became possible to go from an article abstract to the actual observational data used to write the article.

We propose to extend these methods to formalize the links between astronomical resources maintained by different archives, namely the metadata describing astronomical observations and publications. Much of the infrastructure to do this already exists, having been created and curated by groups such as the CDS and the NASA data centers. Our role would be to develop the connection framework by collating and building datasets of properties necessary to interlink observations and publications more tightly. This can only be done with the close cooperation and commitment of our partner organizations.

In modern terms this dense interlinking may be formalized using a Resource Description Framework (RDF) model (W3C, 2009): an object has properties which have values. For example *<a paper> <is written by> <authors>* or *<a paper> <is cited by> <papers>*. Chaining these relationships together and performing functions on the results yields what we call Second Order Operators (Kurtz, 1992). These operations can have very powerful properties. For example, chaining *<a word phrase> <is contained in> <papers>* with *<papers> <are cited by> <papers>* yields a list of papers which cite papers which contain the phrase. Sorting this list by frequency of appearance yields review articles on the original word phrase, which could be anything from "dark energy", to "neutrino mass", or "your favorite subject here." The ADS has internally been making use of the relationships between bibliographic metadata records since its inception to provide some of its more advanced retrieval capabilities.

Observational metadata may be similarly modeled: an observation has an observing proposal, a position in the sky, a time of observation, an instrument, a telescope, an observer, and a P.I., among other properites. These properties will likewise have attributes: the observing proposal will contain words, and may contain a list of observations; the P.I. may have written papers describing such obervations; the position on the sky may correspond with a known object; the instrument will have a type (e.g. imager, spectrograph) and settings (filter, wavelength range, resolution); a combination of time, position and telescope will yield air mass and moon phase, and so on.

A brief example will show the power of such a system. Given a list of the positions of extragalactic globular clusters (for example from NED) one could search for all observations which are IR spectra. Downloading and analyzing the CaII triplet in these spectra would allow one to research the metallicity distribution in the nearby universe. The links from the observing proposals for these spectra to the journal articles, as well as the proposals themselves, would allow the user to determine the original intent of the observations, and whether the proposed work had, indeed, already been done. This new system of observation interlinking would clearly be tightly integrated with the existing infrastructure. For example, the list of observations of IR spectra of extragalactic globular clusters will link to a list of observing proposals, which will link to a set of journal articles which will have been cited by other journal articles, etc.

For this proposal to be implemented three sets of (distributed) metadata resources need to exist and be available in machine-readable format for harvesting and indexing: 1) observing proposals, their attributes and links to observations; 2) observations and their attributes; 3) instruments, their attributes and capabilities. Clearly as the data and nomenclature interoperability standards being developed by the International Virtual Observatory Alliance come into wide acceptance and use, the abilities of the system we propose will be greatly enhanced, especially in terms of automated linking and federation of datasets. Even before this occurs, however, we believe that a dense, structured semantic interlinking of astronomical observations and datasets with the existing system of literature and astronomical objects will have tremendous power and usefulness.

## 2. The Indexing of the Full Text of Astronomy Research Publications

The ability to search the full text of modern astronomy journal articles is the most requested feature for addition into the ADS. Recently a group of librarians from five of the world's largest astronomy libraries signed a joint letter requesting the development of this capability. The ADS has long had a full-text search of the pages it digitized, essentially the pre-1997 astrophysics literature, but it does not have the data to provide a search capability of the modern literature, which is being hosted on the publishers' web sites.

The ADS currently provides some powerful searching capabilities which are based on words in the abstract, authors, references, citations, astronomical objects referenced, or co-readership analysis. However, some types of queries, especially those seeking exact details or facts, can often only be achieved by searching the full-text in the ADS, which excludes the current literature. One important example is to search the contents of articles for the names of telescopes and instruments, object names, and grant numbers. Studies of this sort are commonly used in funding decisions and are currently carried out by having staff manually scan the current literature for relevant keywords (see, e.g., Stevens-Rayburn and Grothkopf 2007), an inefficient and error-prone activity.

Making the full-text of current articles available to the ADS would also enable us to use text mining techniques in order to enrich our existing metadata. This includes, for instance, the extraction of keywords from the text, the identification of entities such as project names, missions, and telescopes, and the extraction of references to other papers and datasets. In addition, it would provide us with the opportunity to apply some of the current state-of-the-art techniques in machine learning to the body of text, with the aim of extracting structured information from it. This includes identifying key concepts, detecting facts and relationships, and discovering correlations between entities. The availability and formalization of these factual data about the papers will improve the recall of searches in our current system and will provide an ideal dataset for some of the applications envisoned in the emerging field of astro-informatics.

## Recommendations

This paper has briefly described the history of the ADS project over the past two decades. Its accomplishments have been a result of its curation efforts, technical expertise and broad support within the astronomical and publishing communities. Thanks to our partnerships, we have been able to create a system which goes beyond a simple search engine but has rather become a model for a community-based portal for scientific research. We believe that this model can be successfully extended more broadly to the network of resources available to the Astronomical Community, through a process of standardization and sharing of metadata related to observations. Many observatories already provide metadata of this sort to the ADS: Chandra, XMM, STScI, and NOAO are examples. Not all do, however. We request that the committee, in support of this and similar efforts **recommend that observatories and archives make available, and permit the harvesting and indexing of metadata concerning publically accessible data and datasets by organizations such as ADS, CDS, NED, and the IVOA members.** This would include observing proposals, observing logs, and FITS headers. We note expressly that we are referring to publically available data. When and whether to make data public is a separate (albeit very important) issue.

The ADS is also uniquely qualified to becoming a "Dark Archive" for astronomical publications currently being published electronically. Under this scenario, the ADS would archive the current content of electronic publications for the purpose of preservation and information discovery. This would ensure longevity to astronomical research papers, and enable search capabilities on the full-text of these publications in a unified environment. Providing the full-text of articles to an outside entity, and allowing it to index them, is an important step for a scholarly society and publisher to make. A few, including the AAS and the RAS, have already entrusted this to the ADS. We request that the committee **recommend that publishers provide the full text of astronomy research articles to the ADS, and allow the ADS to index them for the purpose of providing full-text search capabilities to its users.** We note expressly that this says nothing about the ability of ADS users to access these articles, which would remain, as now, under the control and responsibility of the individual journal publishers.

## Conclusions

Over the past decade organizations such as Google and the Internet Movie Database have grown from tools for the cognoscenti to major economic powerhouses, comparable with automobile manufactures or television networks. These, and many other entities, large and small, may be thought of as parts of the emerging brain of the superorganism which is our now globally interconnected society. This white paper requests that the ASTRO2010 committee recommend the freeing of certain key memory synapses in the collective astronomical intelligence, for the long-term benefit of our community.